\documentstyle[12lomcon,cite,graphicx]{article}

\begin{document}

\title{NEUTRINO SPIN-FLAVOR OSCILLATIONS IN
  \\
  RAPIDLY VARYING MAGNETIC FIELDS}

\author{ M.Dvornikov \footnote{e-mail: maxim.dvornikov@phys.jyu.fi}}

\address{Department of Physics, University of
Jyv\"{a}skyl\"{a}, Finland and \\
IZMIRAN, Troitsk, Russia}


\maketitle\abstracts{ The general formalism for the description of
neutrino oscillations in arbitrary rapidly varying external fields
is elaborated. We obtain the new effective Hamiltonian which
determines the evolution of the averaged neutrino wave function.
The general technique is applied to the neutrino oscillations in
rapidly varying magnetic fields. We evaluate the transition
probabilities of the neutrino spin-flavor oscillations in magnetic
fields of the Sun and compare them with the numerical solutions of
the Schr\"{o}dinger equation with the exact Hamiltonian.}

One of the most interesting problems in the neutrino physics is
the solar neutrino deficit. Nowadays it is experimentally
established that the disappearance of solar electron neutrinos can
be accounted for by the LMA-MSW solution. Apart from the LMA-MSW
solution other theoretical models of the solar neutrino
oscillations like spin-flavor precession (see
work~\cite{AkhPetSmi93} and references therein) are also
considered. We mention that neutrino oscillations in
electromagnetic fields of various configurations were examined in
our previous works (see Refs.~\cite{DvoStuYF}).

In this paper we study neutrino oscillations in presence of
general rapidly varying fields. We derive the new effective
Hamiltonian for the time evolution of the averaged neutrino wave
function. Then we apply the general technique to neutrino
oscillations in rapidly varying magnetic fields and discuss the
neutrino conversion in solar magnetic fields. We present the
numerical solutions of Schr\"{o}dinger equation for the neutrino
system interacting with the constant transversal and twisting
magnetic fields and compare them with the approximate analytical
solutions found in this work.

Let us consider the evolution of the two neutrinos $\nu=(\nu_1,
\nu_2)$, which can belong to different flavors and helicity
states. The evolution of the system is described by the
Schr\"{o}dinger type differential equation,
\begin{equation}\label{shr}
  i\frac{d\nu}{dt}=H\nu.
\end{equation}
Here we do not specify the explicit form of the Hamiltonian but
just suppose that it is decomposed into two terms,
\begin{equation}\label{ham}
  H=H_0+{\cal H},
  \quad
  {\cal H}(t+T)={\cal H}(t).
\end{equation}
The Hamiltonian $H_0$ corresponds to the neutrino interaction with
constant or slowly varying external fields. In presence of only
this term the solution of Eq.~(\ref{shr}) can be easily found. It
is known to be periodical with the typical frequency $\Omega_0\sim
1/L_{\rm eff}$, where $L_{\rm eff}$ is the oscillations length.
The function ${\cal H}(t)$ corresponds to rapidly varying external
fields. The frequency $\omega=2\pi/T$ should be much greater than
$\Omega_0$.

We will seek the solution of Eqs.~(\ref{shr}) and (\ref{ham}) in
the form (see also Ref.~\cite{Kap51}),
\begin{equation}\label{formsol}
  \nu(t)=\nu_0(t)+\xi(t).
\end{equation}
In Eq.~(\ref{formsol}) the function $\xi(t)$ is the small rapidly
oscillating one with zero mean value and the function $\nu_0(t)$
is the slowly varying one during $T$. The mean value of a function
is the time averaging over the period $T$. After substituting
Eq.~(\ref{formsol}) in Eq.~(\ref{shr}) we obtain two groups of
rapidly and slowly varying terms in Eq.~(\ref{shr}). In order for
this expression to be an identity the terms of each group in
left-handed side of the equation should be equal to the terms of
the same group in right-handed side. With help of this fact one
finds the time dependence of $\xi$ in the explicit form,
\begin{equation}\label{xisol}
  \xi(t)=-i
  \left(
    \int{\cal H}(t)\ {\rm d}t
  \right)
  \nu_0(t).
\end{equation}
Averaging residual terms in Eq.~(\ref{shr}) over the period $T$
and accounting for Eq.~(\ref{xisol}) we obtain the equation for
the function $\nu_0$,
\begin{equation}\label{eqnu0}
  i\frac{d\nu_0}{dt}=H_{\rm eff}\nu_0,
  \quad
  H_{\rm eff}=H_0-i
  \overline{{\cal H}
  \left(
    \smallint{\cal H}{\rm d}t
  \right)}.
\end{equation}
In derivation of Eq.~(\ref{eqnu0}) we do not make any assumptions
about the smallness of the interaction described by the
Hamiltonian ${\cal H}$.

On the basis of the general technique we discuss the neutrino
evolution in matter under the influence of a combination of the
constant transversal ${\bf B}_0$ and twisting ${\bf B}({\bf r})$
magnetic fields. In our case the Hamiltonians $H_0$ and ${\cal H}$
have the form
\begin{equation}\label{B0ham}
  H_0=
  \left(
    \begin{array}{cc}
      V/2 & \mu B_0 \\
      \mu B_0 & -V/2 \
    \end{array}
  \right),
  \quad
  {\cal H}=
  \left(
    \begin{array}{cc}
      0 & \mu B e^{-i\omega t} \\
      \mu B e^{i\omega t} & 0 \
    \end{array}
  \right),
\end{equation}
where $\mu$ is the neutrino magnetic moment. In Eq.~(\ref{B0ham})
we introduce the quantity $V/2=(\Delta m^2/4E)\Theta-G_F n_{\rm
eff}/\sqrt{2}$, where $\Theta$ is the function of the vacuum
mixing angle $\theta_{\rm vac}$ (see Ref.~\cite{LikStu95JETPeng}),
$\Delta m^2$ is the difference of the neutrino mass squared, $E$
is the neutrino energy, $n_{\rm eff}$ is the effective matter
density, $\omega$ is the frequency of the transversal magnetic
field variation, $G_F$ is the Fermi constant.

Using Eqs.~(\ref{eqnu0})-(\ref{B0ham}) it is possible to derive
the expression for $H_{\rm eff}$,
\begin{equation}\label{Bhameff}
  H_{\rm eff}=
  \left(
    \begin{array}{cc}
      V/2-(\mu B)^2/\omega & \mu B_0 \\
      \mu B_0 & -V/2+(\mu B)^2/\omega \
    \end{array}
  \right).
\end{equation}
Since $B_0$ and $B$ do not depend on time in Eq.~(\ref{Bhameff}),
we can solve Eq.~(\ref{eqnu0}) for the effective Hamiltonian given
in Eq.~(\ref{Bhameff}). The transition probability is expressed in
the following way
\begin{equation}\label{transprob}
  P(t)=A\sin^2
  \left(
    \frac{\pi t}{L}
  \right),
\end{equation}
where
\begin{equation}\label{A}
  A=\frac{(\mu B_0)^2}
  {
  \left[
    V/2-(\mu B)^2/\omega
  \right]^2+
  (\mu B_0)^2},
  \quad
  \frac{\pi}{L}=
  \sqrt{
  \left[
    V/2-(\mu B)^2/\omega
  \right]^2+
  (\mu B_0)^2}.
\end{equation}
As it results from Eqs.~(\ref{transprob}) and (\ref{A}), if the
following condition is satisfied,
\begin{equation}\label{res}
  \frac{V}{2}\simeq\frac{(\mu B)^2}{\omega},
\end{equation}
then $A\simeq 1$ and the transition probability can achieve great
values. This phenomenon is analogous to resonance amplification of
spin-flavor oscillations.

It is interesting to discuss the situation when $B\gg B_0$. In
this case the value of $A_0=A(B=0)$ is much less than unity. Hence
the transition probability at the absence of the additional
twisting magnetic field is small. In this case, if we choose the
parameters according to Eq.~(\ref{res}), the amplitude of the
transition probability becomes great. From Eqs.~(\ref{A}) and
(\ref{res}) one can derive the restriction imposed on $B$ and
$B_0$, $({B}/{B_0})^2\gg 1$.

Now we consider one of the possible applications of the developed
technique to neutrino spin-flavor oscillations in magnetic fields
of the Sun. The solar magnetic field is unlikely to be only either
constant transversal or twisting. Therefore we can apply the
method elaborated in this paper to the description of the solar
neutrino conversion. We examine one of the possible channels of
neutrino oscillations, namely $\nu_{e L}\leftrightarrow\nu_{\mu
R}$ conversion. First one should estimate the parameter $V$ in
Eq.~(\ref{B0ham}). In our case $\Theta=(1+\cos 2\theta_{\rm
vac})/2$ and the effective matter density $n_{\rm
eff}=(n_e-n_n/2)$. Let us discuss the neutrino with the following
properties, $\Delta m^2\sim 10^{-5}\ {\rm eV}^2$, $\theta_{\rm
vac}\sim \pi/4$. We take the neutrino energy $E\sim 10\ {\rm
MeV}$. Matter is supposed to consist mainly of hydrogen and has
the density $d\sim 1.4\ {\rm g}/{\rm cm}^{3}$. For these
parameters we obtain that $V/2\sim 10^{-15}\ {\rm eV}$.

We can evaluate the strength of the magnetic fields necessary for
the $10\%$ conversion of the initial $\nu_{e L}$ beam. We assume
that neutrino conversion occurs along the distance $D\simeq
3.2\times 10^{14}\ {\rm eV}^{-1}\sim R_\odot/10$, where $R_\odot$
is the solar radius. Neutrinos are supposed to have the
transitional magnetic moment $\mu=10^{-11}\mu_{\rm B}$. Setting
$P(t)=0.1$ in Eq.~(\ref{transprob}) we obtain $B_0\simeq 17.6\
{\rm kG}$. Then we can derive the strength of the twisting
magnetic field: $B\simeq 3.2 B_0\simeq 56.3\ {\rm kG}$. The
resonance condition~(\ref{res}) is satisfied if $\omega\simeq
1.0\times 10^{-14}\ {\rm eV}$.

\begin{figure}
  \centering
  \includegraphics{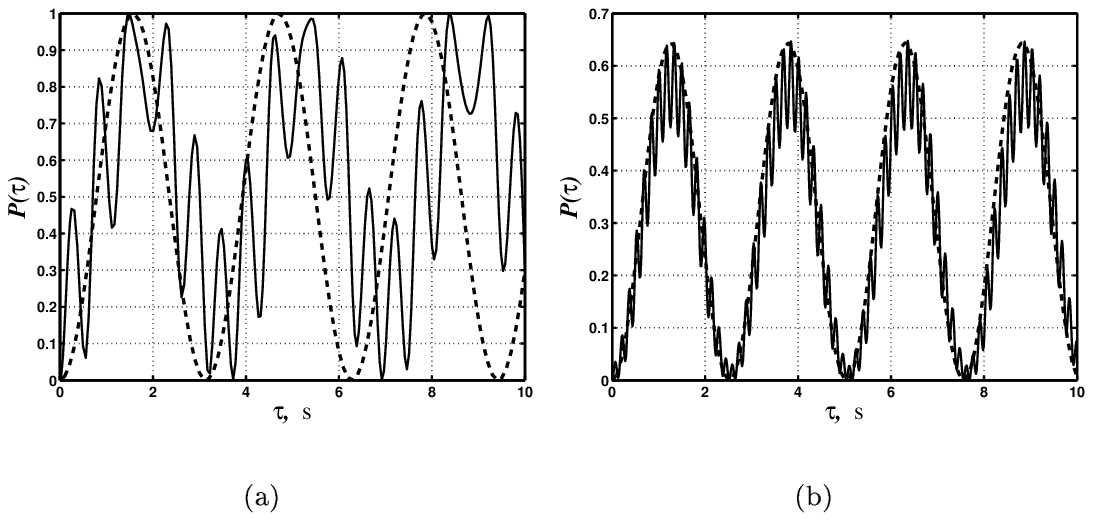}
  \caption{Neutrino transition probability for
  $\nu_{e L}\leftrightarrow\nu_{\mu R}$ oscillations;
  (a) $\omega=1.0\times10^{-14}\ {\rm eV}$;
  (b) $\omega=4.0\times10^{-14}\ {\rm eV}$.}
  \label{numtrpr}
\end{figure}

In order to substantiate the correctness of the approach developed
in our paper we obtain the numerical solutions of Eq.~(\ref{shr})
with the Hamiltonian given in Eq.~(\ref{B0ham}). The probabilities
of neutrino oscillations are shown on Fig.~\ref{numtrpr} versus
$\tau=1.5 t$. We also compare them with the approximate analytical
formula (\ref{transprob}). The solid line corresponds to the
transition probability computed with help of the numerical
solution of Eq.~(\ref{shr}) for the parameters taken above. The
dashed line represents the transition probability given in
Eq.~(\ref{transprob}).
It can be seen in Fig.~\ref{numtrpr} that the numerical expression
for the transition probability approaches to the approximate
formula given in Eq.~(\ref{transprob}) at great frequencies of the
twisting magnetic field. This comparison proves the validity of
the elaborated technique.

\section*{Acknowledgments}

This research was supported by the Academy of Finland under the contract No.~108875 
and by a grant of Russian Science Support Foundation.
The author is indebted to the organizers of the $12^{\rm th}$
Lomonosov Conference on Elementary Particle Physics for inviting
him to participate in this activity.

\end{document}